\documentclass[aps,
prd,
preprint,
eqsecnum,
amsmath,
amssymb
]{revtex4}

\usepackage{hyperref}
\usepackage{color}
\usepackage{xcolor}

\begin{document}

\title{The inverse spatial Laplacian of spherically symmetric spacetimes} 

\author{Karan Fernandes}
 \email{karan12t@bose.res.in}
\author{Amitabha Lahiri}
\email{amitabha@bose.res.in}
\affiliation{
	S. N. Bose National Centre for Basic Sciences,\\
	Block-JD, Sector III, Salt Lake, Kolkata-700106, INDIA.
}

\date{\today}

\begin{abstract}

We derive the inverse spatial Laplacian for static, spherically symmetric 
backgrounds by solving Poisson's equation for a point source. This is different from
the electrostatic Green function, which is defined on the four dimensional static
spacetime, while the equation we consider is defined on the spatial hypersurface 
of such spacetimes. This Green function is relevant in 
the Hamiltonian dynamics of theories defined on spherically symmetric backgrounds, 
and closed form expressions for the solutions we find are absent in the literature. 
We derive an expression in terms of elementary functions for the Schwarzschild
spacetime, and comment on the relation of this solution with the known 
Green function of the spacetime Laplacian operator.
We also find an expression for the Green 
function on the static pure de Sitter space in terms of hypergeometric functions. We
conclude with a discussion of the constraints of the electromagnetic field.

%our result in the quantization of the electromagnetic field on curved backgrounds.}}

\end{abstract}


\maketitle

%%%%%%%%%%%%%%%%%%%%%%%
\section{ Introduction}\label{intro}
%%%%%%%%%%%%%%%%%%%%%%%
Let us consider a four dimensional, static, spherically symmetric background with metric 
\begin{equation}
ds^2  = g_{\alpha\beta} dx^{\alpha} dx^{\beta} 
 = - \lambda^2 dt^2 + h_{\alpha\beta} dx^\alpha dx^\beta\,.
\label{met}
\end{equation}
By static, we mean that the spacetime admits a hypersurface orthogonal timelike Killing 
vector $\xi^\mu\,,$ such that $\xi^\mu \xi_\mu = -\lambda^2\,.$ 
%This background contains spacelike hypersurface $\Sigma$ with an induced metric 
The induced metric on the spacelike hypersurface $\Sigma$ is $h_{\alpha\beta} 
= g_{\alpha\beta} + \lambda^{-2} \xi_\alpha\xi_\beta\,,$ 
and since the hypersurface is assumed to be spherically symmetric, 
we can write $\lambda = \lambda(r)\,.$

The object of our interest in this paper is the Green function 
$\widetilde{G}(\vec{x},\vec{y})$ for the induced spatial Laplacian operator,
 which formally satisfies the equation 
\begin{equation}
%\mathcal{D}_{\mu}\mathcal{D}^{\mu} \Phi(\vec{x}, \vec{y}) = 
%\frac{\delta^3 \left(\vec{x} - \vec{y} \right)}{\sqrt{h}} \equiv 
{D}_{\mu} {D}^{\mu} \widetilde{G}(\vec{x}, \vec{y}) =
- 4 \pi \delta \left( \vec{x},\vec{y} \right) \, ,
\label{IL.eq}
\end{equation}
where $D_\mu$ is the induced covariant derivative compatible with the induced metric,
\begin{equation}
{D}_{\mu} h_{\alpha \beta} = 0 \,,
\end{equation}
and the 3-dimensional covariant delta function $\delta\left(\vec{x}, 
\vec{y} \right)$ is defined by
\begin{equation}
\int\limits_\sigma d^3x \sqrt{\det h(\vec{x})} f(\vec{x})\, 
\delta\left(\vec{x}, \vec{y} \right) = f(\vec{y})\,,
\label{delta.def}
\end{equation}
for all well-behaved functions $f(\vec{x})$ if $\sigma\subseteq\Sigma$ 
includes the point $\vec{y}$, and zero otherwise. 
This Green function is relevant for the Hamiltonian dynamics of fields on curved backgrounds,
as we discuss below.

On the other hand, a different Green function appears in solving for 
the Coulomb potential in static spherically symmetric spacetimes,
and a closed form expression for it is well known.
%%% The study of the Maxwell field on spherically symmetric spacetimes has also led to 
%%% derivation of the static, scalar Green's function of the spacetime Laplacian on these 
%%%  spacetimes. 
For Maxwell's equation 
\begin{equation}
\nabla_{\alpha} F^{\alpha \mu} = -4 \pi J^{\mu} \, ,
\label{max.gen}
\end{equation}
where $F_{\mu \nu} = \partial_{\mu} A_{\nu} - \partial_{\nu} A_{\mu}$ is the usual electromagnetic field strength tensor. 
In what follows, $a_{\mu} = h^{\nu}_{\mu} A_{\nu}$ and $\phi =  \xi^{\alpha}A_{\alpha}$ represent the spatial and temporal components of the spacetime field $A_{\mu}$ respectively.
By defining the electric field as
\begin{equation}
e^{\mu} := \lambda^{-1} \xi_{\alpha}F^{\mu \alpha} \, , 
\end{equation}
we find that the contraction of Eq.\eqref{max.gen} with $\lambda^{-1} \xi_{\mu}$, 
equivalent to setting $\mu=0$, leads to
\begin{equation} 
D_{\mu} e^{\mu} = D_{\mu}\left(\lambda^{-1} D^{\mu} \phi  - \lambda^{-1} \pounds_{\xi} a^{\mu} \right) = -4 \pi J^{0} \, ,
\label{max.con}
\end{equation}
where $J^{0} = \lambda^{-1} \xi_{\mu} J^{\mu}$ and $\pounds_{\xi}$ 
is the Lie derivative with respect to $\xi^{\alpha}$. 
If we also set $\pounds_\xi \phi = 0 = \pounds_\xi a_\mu\,,$ and take a point charge by 
setting $J^{0} = \delta\left(\vec{x},\vec{y}\right)$, Eq.\eqref{max.con} reduces 
to that for the Green function for the electrostatic potential, 
\begin{equation}
%D^x_{\mu} e^{\mu}(\vec{x}) = 
%D^{x}_{\mu}\left(\lambda^{-1}(\vec{x}) D_x^{\mu} \phi(\vec{x}) \right) 
D^{x}_{\mu}\left(\lambda^{-1}(\vec{x}) D_x^{\mu} G(\vec{x}, \vec{y}) \right) 
= -4 \pi \delta(\vec{x},\vec{y}) \, .
\label{max.gff}
\end{equation}
The left hand side of Eq.\eqref{max.gff} is nothing but the the action of 
the d'Alembertian on time-independent functions, for which the expansion
\begin{equation}
\nabla_{\mu} \nabla^{\mu} \phi = \lambda D_{\mu}\left(\lambda^{-1}D^{\mu} 
\phi \right) = D_{\mu} D^{\mu} \phi + \lambda D_{\mu}\left(\lambda^{-1}\right) 
D^{\mu} \phi \, 
\label{max.4d}
\end{equation}
reveals that while Eq.\eqref{IL.eq} and Eq.\eqref{max.gff} are the same in flat space, they differ on curved backgrounds where $\lambda$ is not a constant.
We will call the Green function corresponding to Eq.\eqref{max.gff} the 4d static scalar Green function, and that of Eq.\eqref{IL.eq} the inverse spatial Laplacian. 
%It is also clear that this equation 
%agrees with Eq.\eqref{IL.eq} in flat space, or whenever $\lambda$ is a constant.
%
%On curved spacetimes however, Eq.\eqref{IL.eq} and Eq.\eqref{max.gff} differ. The left hand side
%of Eq.\eqref{max.gff} can be written in coordinates as
%%
%\begin{align}
%	\frac{1}{\lambda \sqrt{h}}\partial_{i} \left( \lambda^{-1} \sqrt{h} h^{i j} \partial_j \Phi({\vec{x}}) \right) 
%	& =  {\lambda}^{-2} \mathcal{D}_i \mathcal{D}^i \Phi({\vec{x}})  + \lambda^{-1} h^{ij}
%	\left(\partial_i \lambda^{-1} \right)  \partial_j \Phi({\vec{x}}) \,,
%	\label{max.4d}
%\end{align}
%%
%where we have written $\Phi = \lambda\phi$\,. The second term is non-zero in general 
%for curved spacetimes. 
%

For the Schwarzschild background, the 4d Green function 
${G(\vec{x}, \vec{y})}$ is known in closed form. It can be 
derived by direct construction of the Hadamard elementary solution~\cite{Cop:1928} and also
using the method of multipole expansion~\cite{J.Math.Phys.12.1845, Hanni:1973fn}. A 
closed form expression was given in~\cite{Linet:1976sq}, which included an additional 
term missed in~\cite{Cop:1928}. This term accounts for the induced 
charge behind the horizon of the black hole on the Schwarzschild background, and
corresponds to the zero mode contribution in the multipole expansion result.
The closed form expression for the static, scalar Green function for the spacetime 
Laplacian on curved backgrounds has found numerous applications ~\cite{Candelas:1980zt, Isoyama:2012in, Frolov:2012ip, Frolov:2012xf, Casals:2013mpa} predominantly 
in its use in determining the self force acting on the particle placed on such 
backgrounds~\cite{Smith:1980tv, Wiseman:2000rm, Barack:1999wf, Beach:2014aba, Frolov:2014gla, Casals:2012qq, Kuchar:2013bla}. Such closed form expressions have additionally 
been determined for the Reissner-N\"ordstrom~\cite{Leaute:1976sn}, and more recently 
for Kerr backgrounds~\cite{Ottewill:2012aj}.

In contrast, the Green function of Eq.\eqref{IL.eq} arises in various contexts which involve fields
on static foliations of spacetime. These include the gravitational 
initial value problem~\cite{Gourgoulhon:2007tn, Dain:2004vi, Pfeiffer:2004qz}, 
metric fluctuations around solutions of the Einstein field equations~\cite{Bardeen:1980kt, 
Antoniadis:2011ib, Mottola:2016mpl}, classical radiation of 
free-falling charges~\cite{Akhmedov:2010ah}, 
%%electromagnetic wave propagation~\cite{Kopeikin:2005jm,Kopeikin:2006gf} 
and more recently, renormalization group equations on curved backgrounds~\cite{Rechenberger:2012dt, Codello:2008vh, Wetterich:2016ewc}, to name a few. This 
Green function is particularly relevant in the context of Hamiltonian 
dynamics of fields. The specific context we have in mind is the constrained dynamics 
of gauge field theories, where this function appears for gravitational~\cite{Deser:1967zzb,Bonazzola:2003dm,Fischer:1996qg} 
and electromagnetic~\cite{physics/9804018, ashtekar:1996bs, igarashi:1998hd, Prescod-Weinstein:2014lua} fields. 
For example, the Maxwell field has the first class Gauss law constraint 
$\mathcal{D}_i \pi^i \approx 0\,,$  which implies the existence of 
redundant or gauge degrees of freedom, which can be eliminated by fixing 
the gauge and then applying Dirac's procedure.  The resultant Dirac brackets 
of the fields and their momenta in the radiation gauge on curved backgrounds 
with horizons involves this Green function~\cite{Fernandes:2016imn}. 
However, while well motivated in the literature, we found no closed form 
expressions for them on curved backgrounds. In this work, we consider 
these functions for spherically symmetric backgrounds, and derive their 
expressions for the Schwarzschild and pure de Sitter cases. For these backgrounds,
the metric of Eq.\eqref{met} takes the form
\begin{align}
	ds^2 %& = g_{\alpha \beta} dx^{\alpha} dx^{\beta} \notag\\
	%% & = - {\lambda(r)}^2 dt^2 + h_{i j} dx^i dx^j \notag\\
	& = - {\lambda(r)}^2 dt^2 + \frac{1}{{\lambda(r)}^2} dr^2 + r^2 d\Omega^2 \,.
	\label{met}
\end{align}
Both backgrounds possess a horizon, defined by $\lambda=0$.

The organization of our paper is as follows. In Sec.~\ref{GF}, we 
review the derivation of the static, scalar Green function for the spacetime 
Laplacian defined on the Schwarzschild background. In 
Sec.~\ref{Sol}, we derive the solutions of Eq.~(\ref{IL.eq}) for the 
Schwarzschild and static pure de Sitter 
backgrounds. While we were able to determine the closed form expression 
for the Schwarzschild case in terms of elementary functions, we were
unable to find a similar expression for the pure de Sitter
background. 
Finally, in Sec.~\ref{Con}, we discuss the relevance 
of our result in the constrained quantization of the Maxwell field 
on spherically symmetric backgrounds.

%%%%%%%%%%%%%%%%%%%%%%%%%%%%%%
\section{Derivation of the 4d static, scalar Green function}\label{GF}
%%%%%%%%%%%%%%%%%%%%%%%%%%%%%%
The Green function corresponding to Eq.\eqref{max.gff} is relevant for Coulomb's law, as we have seen. Let us briefly review 
its derivation on the Schwarzschild 
background following~\cite{J.Math.Phys.12.1845, Hanni:1973fn}, as we will follow a similar procedure 
for deriving the Green function for Eq.\eqref{IL.eq}.

We take $\lambda^2 = g^{rr} = 1 - \frac{2m}{r}$\,, and place a {unit} charge at 
$(r', {\theta}', {\phi}')$.
%
%%\begin{equation}
%%J^{\nu} = (e \delta(r - r')  \delta(\cos \theta - 1) ,0,0,0) \, ,
%%\end{equation}
%
With this choice Eq.\eqref{max.gff} becomes, in explicit coordinates,
\begin{align}
\sin\theta\partial_{r} \left(r^2 \partial_r G \right) + \frac{1}{\left(1 - \frac{2 m}{r} \right)} \partial_{\theta} \left(\sin \theta \partial_{\theta}  G \right) +  \frac{1}{\left(1 - \frac{2 m}{r} \right) \text{sin} \theta} \partial^2_{\phi} G 
\qquad \notag\\
= - 4  \pi \delta(r - r')  \delta(\theta - {\theta}') \delta(\phi - {\phi}') \,,
\label{wgf.me}
\end{align}  
where the delta functions are normalized according to 
\begin{align}
	\int_{2 m}^{\infty} dr \delta(r - r')  = 1\,, \quad \int_{0}^{\pi} d\theta \, \delta (\theta - {\theta}') 
 = 1 \,, \quad \int_{0}^{2 \pi} d \phi \, \delta (\phi - {\phi}')  = 1 \,.
\label{wgf.delta}
\end{align}
Away from the point charge, the right hand side of Eq.\eqref{wgf.me} vanishes, and 
we can expand $G$ as 
\begin{equation}
G(\vec{r}, \vec{r}') = \sum_{l=0}^{\infty} R_l(r, r') P_l(\cos \gamma) \, ,
\label{wgf.sep}
\end{equation}
where $\cos \gamma = \cos \theta \cos {\theta}' + \sin \theta \sin \theta' 
\cos \left( \phi - \phi'\right)$. 
While we could have used the azimuthal symmetry to reduce this to a problem 
in plane polar coordinates $(r, \theta)$\,, the calculations are no more 
complicated for $(r, \theta, \phi)\,,$ so we have chosen to display all coordinates. 
%Spherical symmetry could have been utilized to restrict our consideration to 
%the `$r-\theta$' set of coordinates, in which case the $P_l(\cos \theta)$ 
%functions would be utilized in Eq.\eqref{wgf.sep}. However, the resulting 
%calculations is in no way further complicated by considering the full set of coordinates. 
We note that since $P_l(\cos \gamma)$ is related to the spherical harmonics $Y_{l,m}(\theta, \phi)$ 
via the Legendre addition theorem (cf. Eqs.~(14.30.8), (14.30.9), (14.30.11) of~\cite{NIST})
\begin{equation}
\frac{2l+1}{4 \pi} P_l(\cos \gamma) = \sum_{m=-l}^{l} Y_{l,m}(\theta, \phi) 
Y^*_{l,m}(\theta', \phi')\,,
\label{wgf.spher}
\end{equation}
it further satisfies
\begin{align}
\frac{1}{\sin\theta} \partial_{\theta} \left(\sin\theta \partial_{\theta} P_l(\cos \gamma) \right) + \frac{1}{\sin^2 \theta} \partial_{\phi}^2 P_l(\cos \gamma) = -l(l+1) P_l(\cos\gamma) \,, \label{wgf.lap}\\
\int_{-1}^{1} d \cos \theta \int_{0}^{2 \pi} d \phi P_{l'} 
(\cos \gamma) P_{l} (\cos \gamma) = \delta_{ll'} 
\frac{4 \pi}{2 l + 1}\,.
\label{wgf.norm}
\end{align}
Substituting Eq.\eqref{wgf.sep} in Eq.\eqref{wgf.me} away from the source, 
we find that $R_l(r)$ is a linear combination of two independent solutions, 
\begin{equation}
R_l(r, r') = A_l(r') g_l(r) + B_l(r') f_l(r) \,,
\label{wgf.rsol}
\end{equation}
where $g_l(r)$ and $f_l(r)$ are given by~\cite{J.Math.Phys.12.1845,Israel:1967za}
\begin{align}
g_l(r)  & = \begin{cases} 
 1 \qquad  \qquad & \qquad (\text{for} \,  l = 0) \, , \\
 \frac{2^l l!\, (l-1)!\, m^l}{(2l)!} (r-2 m) \frac{d}{dr}  P_l 
\left(\frac{r}{m} - 1\right)  \,  & \qquad (\text{for} \,  l \neq 0) \,,
\end{cases} 
\label{wgf.wgsol}\\ 
f_l(r) & = - \frac{ (2 l + 1)!}{2^l (l+1)!\, l!\, m^{l+1}}  (r-2 m) 
\frac{d}{dr}  Q_l \left(\frac{r}{m} - 1\right) \,. \qquad \qquad \qquad \quad
\label{wgf.wsol}
\end{align}
Here $P_l$ and $Q_l$ are the Legendre functions of the first and second kind, 
respectively. 
With the exception of $g_0(r) = 1$, the leading term of $g_l(r)$  
is proportional to $r^l$ and diverges as $r \to \infty$. Thus this solution 
is ruled out for large values of $r$. Both $g_l(r)$ and $f_l(r)$ are well 
behaved at the horizon $r = 2m$. However,  $\frac{d}{dr} f_l(r)$ diverges 
logarithmically as $r \to 2m$\,, except when $l=0$. On 
the other hand, the leading behaviour of $f_l(r)$ for large $r$ is proportional to 
$r^{-l-1}$\,.

We can thus write Eq.\eqref{wgf.sep} as
\begin{equation}
G(\vec{r}, \vec{r}') = \begin{cases}
\displaystyle{\sum_{l=0}^{\infty} A_l(r') f_l(r) P_l(\cos \gamma) \qquad \qquad r > r' } \vspace{0.5em} \\ 
\displaystyle{\sum_{l=0}^{\infty} B_l(r') g_l(r) P_l(\cos \gamma) \qquad \qquad r < r'}
\end{cases}
\end{equation}
The continuity of $G$ and discontinuity of $\vec{\nabla}G$  at $\vec{r}=\vec{r}'$ 
%%results in a single set of coefficients $C_l$ for the solution in Eq.\eqref{wgf.me}.
%
%\begin{align}
%\tilde{\Phi}(r,\theta) = \begin{cases}  \displaystyle{\sum_{l=0}^{\infty} C_l g_l(r')f_l(r) P_l(\cos \theta) \qquad \qquad r > r'}  \vspace{0.5em} \\
% \displaystyle{\sum_{l=0}^{\infty} C_l f_l(r') g_l(r) P_l(\cos \theta) \qquad \qquad r < r' \, .}\end{cases}
%\label{wgf.gs}
%\end{align}
%
tells us that by defining $r_< = \min(r\,,r')$ and $r_> = \max(r\,,r')$, 
we can write $G(\vec{r}, \vec{r}')$ as
\begin{equation}
G(\vec{r}_<,\vec{r}_>) = \sum_{l=0}^{\infty} g_l(r_<) f_l(r_>) P_l(\cos \gamma) \,.
\label{wgf.res}
\end{equation}
A bit of algebra now shows that these solutions can be rewritten in the form 
\begin{align}
&G(\vec{r},\vec{r}') = \notag \\
&\frac{1}{r r'} \left[\frac{ (r - m) (r' -m) - m^2 \cos \gamma}{\sqrt{(r - m)^2 + (r' - m)^2 - 2 (r - m)(r' - m)\cos \gamma - m^2 \sin^2 \gamma}} + m \right] \,.
\label{wgf.lin}
\end{align}
This expression, found in~\cite{Linet:1976sq}, differs from a solution 
provided many years 
earlier~\cite{Cop:1928}  because of the term $\frac{m}{r r'}\,,$
which accounts for the zero-mode contribution in Eq.\eqref{wgf.res}. 
The result in Eq.\eqref{wgf.lin} has been derived recently using the heat kernel method and bi-conformal symmetry in~\cite{Frolov:2014kia}.

%%%%%%%%%%%%%%%%%%%%%%%%%%%%%%
\section{Inverse Spatial Laplacian of the Schwarzschild background}\label{Sol}
%%%%%%%%%%%%%%%%%%%%%%%%%%%%%%
Now let us get back to the solution of Eq.\eqref{IL.eq} in the 
Schwarzschild background. With the source at $(r', {\theta}', {\phi}')$ as before,
%%% i.e. we will assume the usual Schwarzschild coordinates. In this case, 
Eq.\eqref{IL.eq} takes the form
\begin{align}
\sin \theta \partial_r \left(r^2 \sqrt{1 - \frac{2 m}{r}} 
\partial_r \widetilde{G} \right) &+ \frac{1}{\sqrt{1 - \frac{2 m}{r}}} 
\partial_{\theta} \left( \sin \theta \partial_{\theta} \widetilde{G} \right) 
+ \frac{1}{\sqrt{1 - \frac{2 m}{r}} \sin \theta} \partial_{\phi}^2 \widetilde{G} 
\notag \\
&= - 4 \pi \delta(r-r') \delta(\theta -\theta') \delta(\phi -\phi') 
\, .
\label{sgf.me}
\end{align}
It will be convenient to make a change of variables from $r$ to 
$y = \frac{r}{m} -1$\,.
%%%, for the purpose of solving this problem. 
%%%After solving the problem, we can revert to the original 
%%%coordinates to express the solution to the above equation. 
After we find the solution, we can change variables again to 
express the Green function in terms of the original coordinates.

In terms of $y$\,, Eq.\eqref{sgf.me} takes the form 
\begin{align}
\sin \theta  \left[ \partial_y \left((y+1)^2 \sqrt{\frac{y-1}{y+1}} 
\partial_y \widetilde{G} \right)  \right.&
\left.
% \right. \notag\\ \qquad \left. 
 + \sqrt{\frac{y+1}{y-1}} \left(\frac{1}
{ \sin \theta} \partial_{\theta} \left( \sin \theta \partial_{\theta} 
\widetilde{G} \right) + \frac{1}{\sin^2 \theta} \partial_{\phi}^2 \widetilde{G} \right) \right] 
\notag\\ 
&\qquad \qquad = - 4 \pi \frac{\delta(y-y')}{m} \delta(\theta -\theta') \delta(\phi -\phi') 
\, ,
\label{sgf.yme}
\end{align}
with the point source located at $(y',\theta',\phi')$ in the new coordinates.

The angular delta functions satisfy the expressions in Eq.\eqref{wgf.delta}, while the $y$ delta function now satisfies
\begin{equation}
\int_{1}^{\infty} dy \, \delta (y - y')  = 1 \, .
\label{sgf.dfn}
\end{equation}

The first step in our derivation is to consider Eq.\eqref{sgf.yme} far removed from the source. Thus we need to solve the following equation
\begin{equation}
0 = \sqrt{\frac{y-1}{y+1}} \partial_y \left((y+1)^2 \sqrt{\frac{y-1}{y+1}} \partial_y\widetilde{G} \right) + \frac{1}{\sin \theta} \partial_{\theta} \left( \sin \theta \partial_{\theta} \widetilde{G} \right) + \frac{1}{ \sin^2 \theta} \partial_{\phi}^2 \widetilde{G}  \,.
\label{sgf.hom}
\end{equation}
Writing 
\begin{equation}
\widetilde{G} (\vec{y},\vec{y}') = \sum_{l=0}^{\infty} R_l(y,y') P_l(\cos \gamma)\,,
\label{sgf.sep}
\end{equation} 
and substituting Eq.\eqref{sgf.sep} in 
Eq.\eqref{sgf.hom}, we get the differential equation
\begin{equation}
\sqrt{\frac{y-1}{y+1}}\frac{d}{dy} \left((y+1)^2 \sqrt{\frac{y-1}{y+1}} \frac{d}{dy} R_l(y,y') \right) -l(l+1) R_l(y,y') =0 \,.
\label{sgf.Req}
\end{equation}
%
%The general solution of Eq.\eqref{sgf.Req} can be derived by employing the ansatz $R_l(y) = P_{\mu}^{\nu}(y) A(y)$, the details of which are provided in (\ref{A}. 

We have described the solution of Eq.\eqref{sgf.Req} in Appendix~\ref{App}.
The general solution is given in Eq.\eqref{app.gensol}, and it is of the form
\begin{equation}
R_l(y,y') = A_l(y') g_l(y) + B_l(y') f_l(y)\,,
\label{sgf.sol}
\end{equation} 
where the functions $g_l(y)$ and $f_l(y)$ involve Legendre polynomials of 
fractional degree, with the argument $y>1$. Legendre polynomials of fractional 
degree can be described in terms of hypergeometric functions, for which there 
exist several representations. A particular representation which we will use 
is (cf. pp 153-163, Table entry 10 and 28, of~\cite{MO:1966})
\begin{align}
P_{\nu}^{\mu}(y) & = \frac{\Gamma \left(-\nu - \frac{1}{2} \right)}
{2^{\nu + 1} \sqrt{\pi} \Gamma \left(-\nu - \mu \right)} y^{-\nu + \mu - 1} 
(y^2 - 1)^{- \frac{\mu}{2}}\times \notag \\
& \qquad\qquad  \times \,_2 F_1 \left(\frac{1 + \nu - \mu}{2}, 
\frac{2 + \nu - \mu}{2}; \nu + \frac{3}{2}; \frac{1}{y^2} \right)  \notag \\
& \qquad\qquad + \frac{2^{\nu} \Gamma \left(\nu + \frac{1}{2} \right)}{ \sqrt{\pi} 
	\Gamma \left(1 + \nu - \mu \right)} y^{\nu + \mu} (y^2 - 1)^{- \frac{\mu}{2}}  \times
\notag \\ 
& \qquad \qquad
\,_2 F_1 \left(\frac{-\nu - \mu}{2}, \frac{1 -\nu - \mu}{2}; -\nu + \frac{1}{2}; 
\frac{1}{y^2} \right) \, , \notag \\
e^{-i \pi \mu} Q_{\nu}^{\mu}(y) & = \frac{\sqrt{\pi} 
	\Gamma\left( 1 + \nu + \mu \right)}{2^{\nu + 1} \Gamma \left(\frac{3}{2} 
	+ \nu \right)} y^{-\nu - \mu - 1} (y^2 - 1)^{\frac{\mu}{2}} \times \notag \\
& \qquad \qquad \times \,
_2 F_1 \left(\frac{\nu + \mu + 2}{2} , \frac{\nu+ \mu +1}{2}; \nu + \frac{3}{2}; 
\frac{1}{y^2} \right) \, ,
\label{sgf.ghgf}
\end{align}
The solutions $g_l(y)$ and $f_l(y)$ make use of these solutions for the case 
of $\mu = \frac{1}{2}$ and $\nu = l$\, as shown in Eq.\eqref{app.gensol} of 
Appendix A, and can be written as
\begin{align}
g_l(y) & = \frac{1}{\sqrt{y+1}} \left[\frac{1}{2^{l + 1}} y^{-l - \frac{1}{2}}  \,_2 F_1 \left(\frac{l + \frac{1}{2}}{2}, \frac{l+ \frac{3}{2}}{2}; l + \frac{3}{2}; \frac{1}{y^2} \right) \right. \notag \\
& \left. \qquad + 2^{l} y^{l+ \frac{1}{2}}  \,_2 F_1 \left(\frac{-l - \frac{1}{2}}{2}, \frac{-l + \frac{1}{2}}{2}; -l + \frac{1}{2}; \frac{1}{y^2} \right) \right] \, , \notag \\
f_l(y) & = \sqrt{y - 1} \left[\frac{1}{2^{l}} y^{-l - \frac{3}{2}} \,_2 F_1 \left(\frac{l + \frac{5}{2}}{2} , \frac{l + \frac{3}{2}}{2}; l + \frac{3}{2}; \frac{1}{y^2} \right) \right] \, .
\label{sgf.main}
\end{align}

It turns out that the functions given in Eq.\eqref{sgf.main} admit expressions 
in terms of more elementary functions, which we will now describe. These 
expressions will be relevant in determining the final form  of
the Green function for the spatial Laplacian. The hypergeometric 
functions contained in $g_l(y)$ in Eq.\eqref{sgf.main} are both of the 
following generic form, with the known representation
\begin{equation}
_2 F_1 \left( a , a+ \frac{1}{2}, 2 a + 1, \frac{1}{y^2} \right) = 
2^{2a} \left( \frac{ y + \sqrt{y^2 -1}}{y} \right)^{-2 a}\,,
\end{equation}
where $a$ stands for both $\frac{ l + \frac{1}{2}}{2}$ and 
$\frac{-l -\frac{1}{2}}{2}$ in the above expression. 
We can thus write the expression for $g_l(y)$ as 
\begin{equation}
g_l(y) 
= \frac{1}{\sqrt{2}\sqrt{y+1}} 
{\left[ \left(y + \sqrt{y^2 - 1} \right)^{-l - \frac{1}{2}} + \left(y + \sqrt{y^2 - 1} \right)^{l + \frac{1}{2}} \right]}\,.
%=  \frac{ \left(y + \sqrt{y^2 - 1} \right)^{-l - \frac{1}{2}} + \left(y + \sqrt{y^2 - 1} \right)^{l + \frac{1}{2}}}{ \sqrt{2}\sqrt{y+1}}
\label{sgf.geq}
\end{equation}
Likewise, the hypergeometric function given in $f_l(y)$ 
has the following expression in terms of elementary functions,
\begin{equation}
_2 F_1 \left( b , b+ \frac{1}{2}, 2 b, \frac{1}{y^2} \right) = \frac{2^{2b - 1} y^{2 b} }{\sqrt{y^2 -1}} \left( y + \sqrt{y^2 -1} \right)^{-2 b + 1}\,,
\end{equation}
where $b = \frac{l + \frac{3}{2}}{2}$\,. 
We can thus write $f_l(y)$ as
\begin{equation}
f_l(y) = \sqrt{2} \, \frac{\left(  y + \sqrt{y^2 -1} \right)^{-l - \frac{1}{2}}}{\sqrt{y + 1}}  \, .
\label{sgf.feq}
\end{equation}

The calculation below will require the Wronskian 
of the solutions given in Eq.\eqref{sgf.main}. Using the above expressions, we
readily find that the Wronskian $W(g_l(y),f_l(y),y)$ is given by
\begin{equation}
W(g_l(y),f_l(y),y) = - \frac{(2 l + 1)}{(1+y)^{\frac{3}{2}} \sqrt{y - 1}}\,.
\label{sgf.wro}
\end{equation}
There are two limits to consider of the solutions given in Eq.\eqref{sgf.geq} 
and Eq.\eqref{sgf.feq}, and their derivatives. These are the $y \to 1$ and 
$y \to \infty$ limits, which correspond to $r \to 2 m$ and $r \to \infty$ 
respectively. Before describing these, we note that $g_0(y)$ is a special case 
in that it is a constant, $g_0(y) = 1$ for all values of $y$. 

For all the other terms we find the following. As  $y \to 1$, both $g_l(y) \to 1$ 
and $f_l(y) \to 1$ for all values of $l$, i.e. they are both finite. However, 
all derivatives of $ f_l(y)$ diverge as $y \to 1$, 
while $\frac{d}{dy} g_l(y) \to l(l+1)$ as $y \to 1$. Thus
the near horizon solution must only contain $g_l(y)$, 
and we must set $B_l(y') = 0$ in Eq.\eqref{sgf.sol} in the region between 
$(y',\theta',\phi')$ and the event horizon of the black hole.

On the other hand, as $y \to \infty$ , we find that $f_l(y) \to 0$ for all 
values of $l$, and the derivatives of $f_l(y)$ are also well behaved, but
$g_l(y)$ diverges for $l \neq 0$.  We must thus set $A_l(y') = 0$ in Eq.\eqref{sgf.sol} 
to describe the region from $(y',\theta',\phi')$ to $\infty$. 

We can therefore write the solution in the 
following way in the two regions,
\begin{equation}
\widetilde{G} \left(\vec{y},\vec{y}'\right) = \begin{cases} \displaystyle{\sum_{l=0}^{\infty} 
A_l(y') g_l(y) P_l(\cos \gamma)\,, \qquad \qquad (y < y')}  \\
\displaystyle{ \sum_{l=0}^{\infty} B_l(y') f_l(y) P_l(\cos \gamma)\,. \qquad \qquad (y > y')}
\end{cases}\end{equation}
Continuity of $\widetilde{G}$ at $y=y'$ implies that $A_l(y') g_l(y') = B_l(y') f_l(y')\,.$ 
Then we can define a constant $C_l$ such that
\begin{equation}
C_l = \frac{A_l(y')}{f_l(y')} = \frac{B_l(y')}{g_l(y')} \, ,
\end{equation}
using which we can write the solution in the form
\begin{equation}
\widetilde{G} \left(\vec{y},\vec{y}'\right)= \begin{cases} \displaystyle{\sum_{l=0}^{\infty} 
C_l f_l(y') g_l(y) P_l(\cos \gamma)\,, \qquad \qquad (y < y')}  \\
\displaystyle{ \sum_{l=0}^{\infty} C_l g_l(y') f_l(y) P_l(\cos \gamma)\,. \qquad \qquad (y > y')}
\end{cases}
\label{sgf.ssol}
\end{equation}
We can now determine the constants $C_l$ by appropriately integrating Eq.\eqref{sgf.yme}. 
To begin with, we insert Eq.\eqref{sgf.sep} into Eq.\eqref{sgf.yme}, multiply both sides  
with $P_{l'}(\cos \gamma)$ and integrate with respect to $\theta$ and $\phi$ to find 
\begin{equation}
\frac{1}{2 l + 1} \left[\frac{d}{dy} \left((y+1)^2 \sqrt{\frac{y-1}{y+1}} 
\frac{d}{dy} R_l(y) \right) - l(l+1) \sqrt{\frac{y+1}{y-1}} R_l(y) \right] = 
 - \frac{\delta(y-y')}{m}\,.
\label{sgf.int1}
\end{equation}
%\warning{There is a mistake in the referee report. The identity within pt. 17 is actually
%$\displaystyle{\sum_{l=0}^{\infty} \sum_{m=-l}^{l} Y_{l,m}(\theta,\phi) Y^*_{l,m}(\theta',\phi') = \delta(\theta - \theta') \delta(\phi - \phi') = \sum_{l=0}^{\infty} \frac{2l+1}{4 \pi} P_l(\cos \gamma)}$. The absence of the factor $2l+1$ in Eq.\eqref{sgf.ssol} is the reason we can't just integrate with respect to $\theta$ and $\phi$, but rather have to first multiply by $P_{l'}(\cos \gamma)$.}
%
%which is not so straightforward to show, but which can be checked case by case for $l = 0,1,2,..$. \\
%
Integrating Eq.\eqref{sgf.int1} over an infinitesimal region from 
$y' - \epsilon$ to $y' + \epsilon$\,, we get
\begin{align}
- \frac{1}{m} & = \frac{1}{2 l + 1} C_l (y'+1)^2 \sqrt{\frac{y'-1}{y'+1}}\left[g_l(y')
\left.\frac{d f_l(y)}{dy}\right\vert_{y' + \epsilon} 
- f_l(y') \left.\frac{d g_l(y)}{dy} \right\vert_{y' - \epsilon} \right] \notag \\
& = \frac{1}{2 l + 1} C_l (y'+1)^{\frac{3}{2}} \sqrt{y'-1} W(g_l(y'),f_l(y'),y') \notag \\
& = -  C_l\,,
\label{sgf.con}
\end{align}
where in going from the second to the third equality in Eq.\eqref{sgf.con}, we made use of Eq.\eqref{sgf.wro}. Thus we have determined that $C_l$ is independent of $l$\,,
\begin{equation}
C_l = \frac{1}{m}\,,
\end{equation}
and we can write the solution of Eq.\eqref{sgf.yme} as
\begin{equation}
\widetilde{G}\left(\vec{y}_<\,,\vec{y}_>\right) =  \frac{1}{m} \sum_{l=0}^{\infty} 
g_l(y_<)\, f_l(y_>) P_l(\cos \gamma)\,,
\label{sgf.csol}
\end{equation}
%
%\begin{align}
%\Phi\left(y,\theta,\phi\right) & = - \frac{1}{4 \pi m} \sum_{l=0}^{\infty} f_l(y') g_l(y) P_l(\cos \gamma) \qquad \qquad (y < y') \notag \\
%& = - \frac{1}{4 \pi m} \sum_{l=0}^{\infty} g_l(y') f_l(y) P_l(\cos \gamma) \qquad \qquad (y > y') \, .
%\label{sgf.csol}
%\end{align}
%
where $y_< = \min(y\,,y')$ and $y_> = \max(y\,,y')$\,. Using Eq.\eqref{sgf.feq} and Eq.\eqref{sgf.geq}, we find that the product $g_l(y_<)\, f_l(y_>)$ is given by
\begin{align}
g_l(y_<) f_l(y_>) &= \frac{1}{\sqrt{y_<+1} \sqrt{y_>+1}} 
\left[ \left(\frac{y_< + \sqrt{y_<^2 - 1}}{y_> + 
	\sqrt{y_>^2 - 1}} \right)^{\frac{1}{2} + l} \right. \notag \\
& \qquad + \left.
\left( \left(y_< + \sqrt{y_<^2 - 1} \right) \left(y_> 
+ \sqrt{y_>^2 - 1} \right) \right)^{-l - \frac{1}{2}} \right]
\label{sgf.exp}
\end{align}

For the sake of notational convenience, let us define
\begin{align}
A = y_> + \sqrt{y_>^2 - 1} \,  \qquad & {\rm and} \qquad B = y_< + \sqrt{y_<^2 - 1}\,. 
\end{align}
Using Eq.\eqref{sgf.exp}, and the standard expression for the generating 
function for Legendre polynomials
\begin{equation}
\sum_{l=0}^{\infty} t^l P_l(x) = \frac{1}{\sqrt{1 - 2 x t + t^2}}\,,
\end{equation}
we find that Eq.\eqref{sgf.csol} takes the form
\begin{align}
\widetilde{G}\left(\vec{y}_<\,,\vec{y}_>\right) &= \frac{1}{m} \frac{1}{\sqrt{y_<+1} 
	\sqrt{y_> + 1}}\, \times \notag \\
& \qquad \times\, \left[\frac{\sqrt{A  B}}{\sqrt{A^2 + B^2 - 2 \, 
		A \, B \,\cos \gamma}} +  \frac{\sqrt{A  B}}
{\sqrt{A^2\, B^2 + 1 - 2 \, A \, B \,\cos \gamma}} \right] \, .
\end{align}
To write the solution in terms of Schwarzschild coordinates, we simply make 
the substitution for $y$\,,
%%% With $y_> = \frac{r}{m} -1$ and $y_< = \frac{r'}{m} -1$ we get the following result
and write
\begin{align}
&\widetilde{G}\left(\vec{r}, \vec{r}'\right)  = \notag \\
& \frac{1}{\sqrt{r r'}} \left[\frac{\sqrt{\left(\kappa(r) r - m\right)  \left(\kappa(r') r' - m\right)}}{\sqrt{\left(\kappa(r) r - m\right)^2 + \left(\kappa(r') r' - m\right) ^2 - 2 \, \left(\kappa(r) r - m\right)  \, \left(\kappa(r') r' - m\right)  \,\cos \gamma}} \right. \notag\\
& \left. \qquad +\frac{m \sqrt{\left(\kappa(r) r - m\right)   \left(\kappa(r') r' - m\right) }}{\sqrt{\left(\kappa(r) r - m\right) ^2\, \left(\kappa(r') r' - m\right) ^2 + m^4 - 2 \, m^2 \left(\kappa(r) r - m\right)  \, \left(\kappa(r') r' - m\right)  \,\cos \gamma}} \right] \, ,
%&\Phi\left(\vec{r}_<, \vec{r}_>\right)  = \notag \\
%& \frac{1}{\sqrt{r_< r_>}} \left[\frac{\sqrt{\left(\kappa(r_<) r_< - m\right)  \left(\kappa(r_>) r_> - m\right)}}{\sqrt{\left(\kappa(r_<) r_< - m\right)^2 + \left(\kappa(r_>) r_> - m\right) ^2 - 2 \, \left(\kappa(r_<) r_< - m\right)  \, \left(\kappa(r_>) r_> - m\right)  \,\cos \gamma}} \right. \notag\\
%& \left. \qquad +\frac{m \sqrt{\left(\kappa(r_<) r_< - m\right)   \left(\kappa(r_>) r_> - m\right) }}{\sqrt{\left(\kappa(r_<) r_< - m\right) ^2\, \left(\kappa(r_>) r_> - m\right) ^2 + m^4 - 2 \, m^2 \left(\kappa(r_<) r_< - m\right)  \, \left(\kappa(r_>) r_> - m\right)  \,\cos \gamma}} \right] \, ,
\label{sgf.sgf}
\end{align}
where we have defined $\kappa(r) = 1 + \lambda(r) = 1 + \sqrt{ 1 - \frac{2 m}{r}}$\,, 
and $\kappa(r')$ similarly.  
As noted earlier, we see that as we take the flat space limit ($m \to 0$), this solution 
as well as that of Eq.\eqref{wgf.lin} reduce to the Green function of flat space.
We also note that just as in the Green function 
result given in the previous section, this solution is regular at the horizon.

%%%%%%%%%%%%%%%%%%%%%%%%%%%%%%%%%%%%%%
\section{Inverse spatial Laplacian of the de Sitter background}
%%%%%%%%%%%%%%%%%%%%%%%%%%%%%%%%%%%%%%
We now turn our attention to writing a closed form expression for the Green function 
on a de Sitter background. 
The scalar de Sitter Green function for cosmological 
de Sitter spacetimes has been derived in~\cite{Bunch:1977sq, Chernikov:1968zm, Tagirov:1972vv}. 
In static coordinates, the thermal Green function for the massless scalar field
equation~\cite{Dowker:1977}, as well as the Green function for the massive 
scalar field equation~\cite{Anninos:2011af, Higuchi:1986ww} are known in the 
literature. These Green functions correspond to the de Sitter generalization 
of Eq.\eqref{max.4d}, whereas we will be concerned  with the derivation 
of the solution of the inverse spatial Laplacian, i.e. of Eq.\eqref{IL.eq}. 
The procedure described in this subsection can be used 
for finding the solution of Eq.\eqref{max.gff} as well. 

For pure de Sitter space with cosmological constant $\Lambda\,,$ we have
$\lambda(r)^2 = 1 - \frac{r^2}{L^2}$, where $L = \sqrt{\frac{3}{\Lambda}}$\,,
working in the quadrant of de Sitter space where the time coordinate 
increases into the future. 
We again make a change of coordinates and write
$y = \frac{r}{L}$.  For this choice, Eq.\eqref{IL.eq} takes the form
\begin{align}
\sin \theta \left[ \partial_y \left( y^2 \sqrt{1 - y^2} \partial_y \widetilde{G} \right)  
+ \frac{1}{\sqrt{1 - y^2}} \right.&\left(\frac{1}{ \sin \theta} \partial_{\theta} \right.
\left( \sin \theta \partial_{\theta} \widetilde{G} \right) +
\left.\left. \frac{1}{\sin^2 \theta} 
\partial_{\phi}^2 \widetilde{G} \right) \right] \notag\\
&= - 4 \pi \frac{\delta (y - y')}{L} \delta ( \theta - \theta')\delta (\phi - \phi')\,.  
\label{dgf.yme}
\end{align}
The delta functions for the angular variables satisfy Eq.\eqref{wgf.delta}, but the $y$ delta function now satisfies
\begin{equation*}
\int_{0}^{1} dy \delta(y - y') = 1\,.
\end{equation*}

As in the Schwarzschild case, we begin by solving the above equation far away from the source
\begin{equation}
\sqrt{1 - y^2} \partial_y \left( y^2 \sqrt{1 - y^2} \partial_y \widetilde{G} \right)  + \frac{1}{\sin \theta} \partial_{\theta} \left( \sin \theta \partial_{\theta} \widetilde{G} \right) + \frac{1}{ \sin^2 \theta} \partial_{\phi}^2 \widetilde{G} = 0\,,
\label{dgf.hom}
\end{equation}
with 
\begin{equation}
\widetilde{G}(\vec{y},\vec{y}') = \sum_{l=0}^{\infty} R_l(y,y') P_l(\cos \gamma)\,.
\label{dgf.sep}
\end{equation} 
Substituting Eq.\eqref{dgf.sep} in Eq.\eqref{dgf.hom}, and using 
Eq.\eqref{wgf.lap}, we get the equation
\begin{equation}
\sqrt{1 - y^2} \frac{d}{dy} \left( y^2 \sqrt{1 - y^2} 
\frac{d}{dy} R_l(y,y') \right) - l(l+1) R_l(y,y') = 0\,.
\label{dgf.Req}
\end{equation}
To find the general solution in this case, it will be convenient to express Eq.\eqref{dgf.Req} in terms of $t = \sqrt{1-y^2}$, which results in
\begin{equation}
\sqrt{1 - t^2} \frac{d}{dt} \left( (1 - t^2)^{\frac{3}{2}}
 \frac{d}{dt} R_l(t,t') \right) - l(l+1) R_l(t,t') = 0\,.
\label{dgf.Req2}
\end{equation} 

Using the ansatz $R_l(t,t') = B_l(t') P_{\mu}^{\nu}(t) A(t)$ as before (see Appendix~\ref{App}), we find the following general solution
\begin{equation}
R_l(t,t') = A'_l(t') (1 - t^2)^{-\frac{1}{4}} P_{\frac{1}{2}}^{l + \frac{1}{2}}(t) + B'_l(t') (1 - t^2)^{-\frac{1}{4}} P_{\frac{1}{2}}^{-l - \frac{1}{2}}(t)\,.
\label{dgf.Req3}
\end{equation}
The Legendre polynomials described in Eq.\eqref{dgf.Req3} can be described in terms 
of hypergeometric functions. For Legendre polynomials defined in the region between $-1$ 
and $+1$, we have (cf. p.166 of~\cite{MO:1966})
\begin{equation}
\Gamma(1 - \mu) P_{\nu}^{\mu}(x)  = 2^{\mu} ( 1 - x^2 )^{-\frac{\mu}{2}} 
\, _2 F_1 \left( \frac{1}{2} + \frac{\nu}{2} -  \frac{\mu}{2}, -  \frac{\nu}{2} -  \frac{\mu}{2} ; 1 - \mu ; 1 - x^2 \right)\,.
\label{dgf.iden}
\end{equation}

By using the expressions in Eq.\eqref{dgf.iden}, and writing the results in terms of the variables $y$ by substituting  $1 - t^2 = y^2$, one can find the following general solution
\begin{equation}
R_l(y,y') = A_l(y') g_l(y) + B_l(y') f_l(y) \, ,
\label{dgf.Rgen}
\end{equation}
where $g_l(y)$ and $f_l(y)$ are now given by
\begin{align}
g_l(y) & = y^l \, _2 F_1 \left(\frac{l}{2}, \frac{l}{2} + 1; \frac{3}{2} + l; y^2 \right) \notag \\
f_l(y) & = \frac{1}{y^{l+1}} \, _2 F_1 \left(\frac{-l - 1}{2}, \frac{-l + 1}{2}; \frac{1}{2} - l; y^2 \right)\, .
\label{dgf.Rsol}
\end{align}
Here, $A_l(y')$ and $B_l(y')$ are real coefficients, and the solutions themselves are 
positive and real in the region between 0 and +1. The Wronskian of the two 
solutions given in Eq.\eqref{dgf.Rsol} satisfies the following relation
\begin{equation}
W(g_l(y),f_l(y),y) = - \frac{2 l + 1}{y^2 \sqrt{1 - y^2}} \, .
\label{dgf.wro}
\end{equation}

Unlike in the Schwarzschild case, we were unable to determine a
closed form expression  of the solutions in terms of elementary functions
for arbitrary $l$.
The solutions for specific choices of $l$ however can be easily determined. 
Using the derivative relations satisfied by the hypergeometric functions, we have derived in Appendix~\ref{app2}
the following general form of the $f_l(y)$ solutions 
\begin{align}
%g_l(y) &= 1  & (l = 0) \, , \notag\\
%& = \frac{\sqrt{1 - y^2}}{y^{l+1}} \left[ \alpha(y) {\sin}^{-1} (y) + \beta(y) \right] &  (l \neq 0) \, ,\notag \\
f_l(y) &= \sum_{n = 0}^{\frac{l-1}{2}} \frac{c_n}{y^{2 n + 2}} & (l \,\text{odd}) \, ,\notag\\
 &= \frac{ \sqrt{1 - y^2}}{y} &  (l = 0) \, ,  \notag\\
&= \sqrt{1 - y^2} \sum_{n=1}^{\frac{l}{2}} \frac{c_n}{y^{2 n + 1}} & (l \, \text{even} \, ; l \neq 0) \, ,
\label{dgf.rep} 
\end{align}
where $c_{\frac{l-1}{2}} = 1$ for the odd $l$ case, and $c_{\frac{l}{2}} = 1$ for the even $l$ case.

To proceed further, we need to determine the behaviour of the solution in the limit $y\to 0$ and
$y\to 1$\,. As before, $g_0(y) =1$, which follows from $\ _2 F_1 \left(0, 1; \frac{3}{2}; y^2 \right) = 1$, 
and will not be considered in the following. As $y \to 0$, we can make use of the following 
derivative relation satisfied by the hypergeometric functions
\begin{equation}
\frac{d}{dx} \, _2 F_1 \left( a , b, c, x \right) = \frac{a \,  b}{c} \, _2 F_1 \left( a + 1 , b + 1, c + 1, x \right) \, , 
\label{dgf.hypr}
\end{equation}
as well as $_2 F_1 \left( a , b, c, 0 \right) = 1$, to determine the behaviour 
of the solutions. We see that $g_l(y)$ and its first derivative vanish, while 
$f_l(y)$ and its first derivative diverge for all values of $l$\,, as $y\to 0$\,. 
We must thus set $B_l = 0$ in 
Eq.\eqref{dgf.Rgen} in the region 
where $y$ can vanish, in order to have regular solutions. 

As $y \to 1$, we need to consider the integral representation of the hypergeometric function to demonstrate 
that $g_l(y)$ is finite while its first derivative diverges, for $l \neq 0$. This is shown in Appendix~\ref{app2}. 
The solutions provided in Eq.\eqref{dgf.rep} tell us the following about the $f_l(y)$ solutions 
in the limit $y \to 1$. While $f_l(y)$ remains finite for all $l$, and $f_l(1) = 0$ when $l$ is even, 
the behaviours of the first derivatives differ for even and odd $l$. The first derivative of $f_l(y)$ 
diverges when $l$ is even, and is finite when $l$ is odd. Regularity of the solutions requires 
that in the region where $y\to 1$, we not only set $A_l = 0$ for all $l\neq 0$, 
but also set $B_l = 0$ for even $l$\,.

We can now determine the general solution $\widetilde{G}(\vec{y},\vec{y}')$ for the point 
source located at $(y', \theta', \phi')$. Away from the source the solution is 
given by Eq.\eqref{dgf.sep}. As explained above, in the region $y<y'$ we simply set $B_l(y') = 0$ and 
sum over all $l$. In the region $y>y'$  we set $A_l(y') = 0$ for all $l \neq 0$ and sum over all odd $l$, but we in addition have the $g_0(y) =1$ term which contributes a constant term. 
Thus, we can write
\begin{align}
\widetilde{G}(\vec{y},\vec{y}')  = \begin{cases} \displaystyle{\sum_{l=0}^{\infty} A_l(y') g_l(y) P_l(\cos \gamma)  \quad  \qquad \qquad \qquad \qquad (y < y') \, ,}\\
\displaystyle{A'_0 + \sum_{l=0}^{\infty} B_{2l+1}(y') f_{2l+1}(y) P_{2l+1}(\cos \gamma) \, \,  \qquad (y>y') \, .} \end{cases}
\label{dgf.sep2}
\end{align} 

Finally, we need to match these solutions at $y=y'$. This sets $A_0 = A'_0\,,$ and 
leads us to define the constant $ C_{2l+1} = \frac{A_{2l+1}(y')}{ f_{2l+1}(y')} =
 \frac{B_{2l+1}(y')}{g_{2l+1}(y')}$, 
and we also find that $A_k$ vanishes for even $k (\neq 0)$. Then we can write
\begin{equation}
\widetilde{G}(\vec{y},\vec{y}') = C_0 + R_{2l+1}(y,y') P_{2l+1}(\cos \gamma)\, ,
\end{equation}
where $C_0 \equiv A_0$ is the constant zero-mode contribution, and 
\begin{align}
R_{2l+1}(y,y') =  \begin{cases} \displaystyle{  \sum_{l=0}^{\infty} C_{2l+1}  
	g_{2l+1}(y) f_{2l+1}(y') \, \qquad  (y < y') \, ,} \\
\displaystyle{\sum_{l=0}^{\infty} C_{2l+1}  f_{2l+1}(y) g_{2l+1}(y') \,  \qquad (y>y') \, .}\end{cases}
\label{dgf.ssol}
\end{align} 

Multiplying both sides of Eq.\eqref{dgf.yme} with $P_{2l'+1}(\cos \gamma)$ and 
integrating with respect to $\theta$ and $\phi$, we get
\begin{equation}
- \frac{\delta(y-y')}{L} = \frac{1}{4 l + 3} \left[\frac{d}{dy} \left( y^2 \sqrt{1 - y^2}\frac{d}{dy} R_{2l+1}(y,y') \right) - \frac{(2l+1)(2l+3)}{\sqrt{1 - y^2} } R_{2l+1}(y,y') \right] \, ,
\label{dgf.aint}
\end{equation}
where we have used Eq.\eqref{wgf.norm}.
%
%\begin{align}
%R_{2l+1}(y) & =  C_{2l+1}  g_{2l+1}(y) f_{2l+1}(y')  \, \qquad  (y < y')\,, \notag \\
%& =  C_{2l+1} f_{2l+1}(y) g_{2l+1}(y')  \,  \qquad (y>y')\,.
%\end{align}}}
%
We next integrate over $y$ from $y' - \epsilon$ to $y' + \epsilon$, 
i.e. over an infinitesimal region about the point source, for which we find
\begin{align}
- \frac{1}{L} & = \frac{1}{4 l + 3} C_{2l+1} {y'}^2 \sqrt{1 - {y'}^2} \,\times \notag \\
& \qquad \times \, \left[g_{2l+1}(y') 
\left.\left( \frac{d}{dy} f_{2l+1}(y) \right) \right\vert_{y' + \epsilon} 
- f_{2l+1}(y') \left.\left( \frac{d}{dy} g_{2l+1}(y) \right) \right\vert_{y' - \epsilon} \right] \notag \\
& = \frac{1}{4 l + 3} C_{2l+1} {y'}^2 \sqrt{1 - {y'}^2}\,  W(g_{2l+1}(y'),f_{2l+1}(y'),y') \notag \\
& = - C_{2l+1} \, ,
\label{dgf.con}
\end{align}
where we have made use of Eq.\eqref{dgf.wro} in going from the second to the third 
equality in Eq.\eqref{dgf.con}. 
Using this, we can write the Green function in the de Sitter case as
\begin{equation}
\widetilde{G} \left(\vec{y}_<,\vec{y}_>\right) = \frac{1}{ L} \sum_{l=0}^{\infty} g_{2l+1}(y_<) f_{2l+1}(y_>) P_{2l+1}(\cos \gamma) \, ,
\label{dgf.csol}
\end{equation} 
where $y_< = \text{min}(y,y')$ and $y_> = \text{max}(y,y')$ as before.
Unlike in the Schwarzschild case, we have not been able to write this 
in a simpler form. We can nonetheless substitute for $y$ in Eq.\eqref{dgf.Rsol} 
and use this in Eq.\eqref{dgf.csol}, by writing $y_< = \frac{r_<}{L}$ and 
 $y_> = \frac{r_>}{L}$, to find the solution in terms of $r\,,$
\begin{align}
\widetilde{G}\left(\vec{r}_<, \vec{r}_> \right) = \frac{1}{r_>^2} & \sum_{l=0}^{\infty} \left(\frac{r_<}{r_>}\right)^{2l+1} \, _2 F_1 \left(l + \frac{1}{2}, l + \frac{3}{2}, 2l + \frac{5}{2}, \frac{3 r_<^2}{\Lambda} \right)\, \times \notag\\
& \qquad \qquad \times \,_2 F_1 \left(- l-1, -l, -2l - \frac{1}{2}, \frac{3 r_>^2}{\Lambda} \right) P_{2l+1}(\cos \gamma)\,.
\label{dgf.crsol}
\end{align}

%%%%%%%%%%%%%%%%%%%%%%%%%%%%%%%%%%
\section{Conclusion} \label{Con}
%%%%%%%%%%%%%%%%%%%%%%%%%%%%%%%%%%
In this paper, we have discussed a new class of static, scalar Green functions 
on spherically symmetric spacetimes, those corresponding to the inverse 
spatial Laplacian defined exclusively on the spatial hypersurface of the 
spherically symmetric spacetime. Specifically, we have derived the inverse 
spatial Laplacian in the form of mode solutions for the 
Schwarzschild and pure de Sitter backgrounds.  We have determined the 
closed form expression for Green function on Schwarzschild spacetime in 
terms of elementary functions, and on the pure de Sitter space in 
terms of hypergeometric functions.

As we have mentioned earlier, one of the places where the spatial Laplacian appears 
is in the constrained quantization of Maxwell fields on static spherical symmetric 
spacetimes with horizons. Let us now briefly discuss the role 
of the Green function in that problem; for more details we refer the 
reader to~\cite{Fernandes:2016imn}. 

For the Maxwell field, an important distinction between the its treatment on 
spacetimes with or without horizons is that the Gauss law constraint in the
former case picks up additional surface terms from the horizon,
\begin{equation}
\Omega(\vec{r}) =  - n_\mu  \pi^\mu(\vec{r})\delta (r - r_H) + 
{D}_\mu \pi^\mu(\vec{r}) \,.
\label{con.con}
\end{equation}
Here $\pi^\mu$ are the momenta conjugate to the hypersurface projected field 
$a_\mu\,, r_H$ is the horizon radius, and $n^\mu$ is the outward pointing normal 
on the horizon. 

This is still a first class constraint, and one way of handling it
is to fix a gauge and find the corresponding Dirac brackets. An interesting choice
of gauge fixing function is one that includes a surface term,
\begin{equation}\label{gf}
\Omega_{gf} = D_\mu a^\mu  - n_\mu a^\mu \delta(r - r_H) \,.
\end{equation}
For this choice, the relevant Dirac bracket becomes
\begin{equation}
\left[a_\mu(\vec{r}),\pi^\nu(\vec{y})\right]_{D}  = \delta(\vec{r},\vec{y})
\delta_\mu^\nu - {D}_\mu^r {D}_y^\nu 
\widetilde{G}\left(\vec{r},\vec{y} \right)\,,
\label{con.isl} 
\end{equation}
where $\delta(\vec{r}, \vec{y})$ is defined in Eq.\eqref{delta.def}.

A different gauge choice would produce a different set of brackets, for example the gauge choice $\Omega_{gf} = D_\mu (\lambda a^\mu)$ produces Dirac brackets in which $\widetilde G$ is replaced by $G$ of Eq.\eqref{max.gff} in the second term on the right hand side, and that term also picks up a factor of $\lambda\,.$

Consider the Dirac bracket in the Schwarzschild background. 
When one of the arguments is at the horizon, e.g., in the limit $y \to r_H$, 
we find for the $r - r$ component that
\begin{align}
\left[a_r(\vec{r}),\pi^r(\vec{y})\right]_{D}\Big\vert_{y \to r_H} &= \delta(r,r_H)+  \kappa_H \frac{2r- m(1+ \cos\gamma)}{2 \left(r^2 - mr(1+ \cos\gamma)\right)^{3/2}} \,. \label{db.isl}
\end{align}
For the other gauge choice mentioned above, only the $\delta(r,r_H)$ remains on the
right hand side of the above equation in the limit $y \to r_H$\,.
The Dirac brackets comprise one aspect that enters into the quantization 
of theories. Since the Gauss' law constraint must be respected by physical 
states of the theory, the surface term contained in the constraint on these 
backgrounds will be relevant to states at the horizon. A complete treatment 
of the quantization of the Maxwell field on static, spherically symmetric 
backgrounds with horizons lies outside the scope of the present work. In 
light of the preceding discussion, we can nevertheless expect that the 
inverse spatial Laplacian will affect the quantization of gauge fields 
near the horizon.

Since both the de Sitter and 
Schwarzschild cases admit a mode expansion, where the functions depending 
on $r$ are ultimately associated Legendre polynomials, it seems plausible to 
presume that a similar result would hold for the Schwarzschild-de Sitter background. 
Unfortunately, we have been unable to find a simple transformation for this 
case since the cubic dependence on $r$ in the lapse function $\lambda$ poses a 
significant obstacle to the procedure. From the nature of the equation to solve 
for the Schwarzschild-de Sitter background, it appears that the solution 
for the corresponding Green function will require a different approach from
what was considered here.

\appendix
%%%%%%%%%%%%%%%%%%%%%%%%%%%%%%%%%%
\section{Derivation of the general solution of the homogeneous equations }\label{App}
%%%%%%%%%%%%%%%%%%%%%%%%%%%%%%%%%%
We seek to solve Eq.\eqref{sgf.Req} and Eq.\eqref{dgf.Req2},
which take the general form
\begin{equation} 
(1-y^2)\frac{d^2}{dy^2}R_l(y,y') +  f(y) \frac{d}{dy} R_l(y,y') + g(y)R_l(y,y') - l(l+1)R_l(y,y') = 0\,.
\label{app.eq}
\end{equation}
%
%where primes denote differentiation with respect to the variable `$y$'. 

We will solve this equation, for the cases of Eq.\eqref{sgf.Req} and 
Eq.\eqref{dgf.Req2}  by making use of the ansatz $R_l(y,y') = B_l(y') P_{\nu}^{\mu}(y)A(y)$. 
We first recall that the Legendre polynomial is a solution of the following 
differential equation
\begin{equation}
(1 - y^2) \frac{d^2}{dy^2}P_{\nu}^{\mu}(y) - 2y \frac{d}{dy} P_{\nu}^{\mu}(y) + \left[ \nu(\nu +1) - \frac{{\mu}^2}{1 - y^2} \right]P_{\nu}^{\mu}(y) = 0\,.
\label{app.Leq}
\end{equation}
Expanding Eq.\eqref{sgf.Req}, we find
\begin{equation}
(1-y^2)\frac{d^2}{dy^2}R_l(y,y') - (2y- 1)\frac{d}{dy}R_l(y,y')+  l(l+1) R(y,y') = 0\,.
\end{equation}
Substituting the ansatz and making use of Eq.\eqref{app.Leq}, we get
\begin{align}
A(y) &\left[-\left( \nu(\nu +1) - \frac{{\mu}^2}{1 - y^2} \right) P_{\nu}^{\mu}(y) 
+ \frac{d}{dy} P_{\nu}^{\mu}(y)\right] \notag \\ 
& \qquad \qquad + P_{\nu}^{\mu}(y)\left[(1-y^2)\frac{d^2}{dy^2}A(y) 
- (2y - 1)\frac{d}{dy}A(y) \right] \notag\\
& \qquad \qquad  \qquad + 2(1-y^2)\frac{d}{dy}P_{\nu}^{\mu}(y)\frac{d}{dy}A(y)  
+ l(l+1)P_{\nu}^{\mu}(y)A(y) = 0\,.
\label{app.exp}
\end{align} 
Collecting terms, we have
\begin{align}
\frac{d}{dy}& P_{\nu}^{\mu}(y) \left[2 (1-y^2) \frac{d}{dy}A(y) + A(y) \right] \notag\\
& \quad + P_{\nu}^{\mu}(y) \left[ (1-y^2)\frac{d^2}{dy^2}A(y) - (2y - 1)\frac{d}{dy}A(y) 
\right. \notag \\ & \qquad \qquad \left.
-\left( \nu(\nu +1) - \frac{{\mu}^2}{1 - y^2} - l(l+1) \right) A(y) \right] = 0\,.
\label{app.reo}
\end{align}
We can now explore the simplest possibility which makes Eq.\eqref{app.reo} true, 
namely, that the coefficients of $\frac{d}{dy}P_{\nu}^{\mu}(y)$ and $P_{\nu}^{\mu}(y)$ 
individually vanish. The coefficient is of $\frac{d}{dy}P_{\nu}^{\mu}(y)$ can 
be trivially solved to give the following solution for $A(y)$
\begin{equation}
A(y) = \left(\frac{y-1}{y+1}\right)^{\frac{1}{4}} \,.
\end{equation}
Substituting this solution back in Eq.\eqref{app.reo} gives us the following expression
\begin{equation}
\left( \frac{1}{4} - {\mu}^2 \right)(y-1)^{-\frac{3}{4}}(y+1)^{-\frac{5}{4}} -\left( \nu (\nu +1) - l(l+1) \right) (y-1)^{\frac{1}{4}}(y+1)^{-\frac{1}{4}} = 0 \, .
\label{app.fin}
\end{equation}
Eq.\eqref{app.fin} holds, provided $\mu = \frac{1}{2}$ and $\nu = l$.

One solution of Eq.\eqref{app.exp} is thus $\left(\frac{y-1}{y+1}\right)^{\frac{1}{4}} P_l^{\frac{1}{2}}(y)$. Since our procedure made use of the Legendre polynomials, we would get another solution by simply using $R_l(y,y') = B_l(y')Q_{\nu}^{\mu}(y) A(y)$, with the same solution for $A(y)$. The general solution is thus found to be 
\begin{equation}
R_l(y,y') = A_l(y') \left(\frac{y-1}{y+1}\right)^{\frac{1}{4}} 
P_l^{\frac{1}{2}}\left(y \right) + B_l(y') \left(\frac{y-1}{y+1}\right)^{\frac{1}{4}} 
\left( i Q_l^{\frac{1}{2}}\left(y\right) \right)\,.
\label{app.gensol}
\end{equation}
Eq.\eqref{sgf.Req} is written as it is since $i \, Q_l^{\frac{1}{2}}\left(y\right)$ is a real solution. 

This procedure can similarly be used in Eq.\eqref{dgf.Req2}, which can be written as
\begin{equation}
(1-t^2)\frac{d^2}{dt^2} R_l(t,t') - 3 t\frac{d}{dt} R_l(t,t') - \frac{l(l+1)}{(1 - t^2)} R_l(t,t') = 0 \, .
\end{equation}
Substitution of the ansatz $R_l(t,t') = B_l(t') P_{\nu}^{\mu}(t) A(t)$ now leads to the following equation
\begin{align}
\frac{d}{dt}P_{\nu}^{\mu}(t) & \left[2 \frac{d}{dt}A(t) (1 - t^2) - A(t) t \right] \notag \\
& \qquad + P_{\nu}^{\mu}(t) \left[(1 - t^2) \frac{d^2}{dt^2}A(t) - 3 t \frac{d}{dt}A(t) 
\right. \notag \\
& \qquad \qquad - \left. \left( \nu (\nu +1) - \frac{{\mu}^2}{1-t^2} + 
\frac{l( l + 1)}{1-t^2} \right) A(t) \right] = 0 \, .
\label{app.ts}
\end{align}
As before, we assume the possibility that the coefficients of the $P_{\nu}^{\mu}(t)$ and $\frac{d}{dt}P_{\nu}^{\mu}(t)$ separately vanish. The coefficient of the latter term vanishing leads to the following simple result for A(t) 
\begin{equation}
A(t) = (1 - t^2)^{- \frac{1}{4}} \, .
\end{equation} 
Substituting this equation back into Eq.\eqref{app.ts} leads to the following result
\begin{equation}
- \left[\frac{3}{4} - \nu (\nu +1) \right]t^2 - \left[ \nu (\nu +1) - \frac{1}{2} + l(l+1) - {\mu}^2 \right]  = 0 \, ,
\end{equation}
which is satisfied for the choice of $\nu =\frac{1}{2}$ and $\mu = l + \frac{1}{2}$.
Since in this case $\nu \pm \mu$ is an integer but $\mu$ is not, the other independent 
solution is not $Q^{\mu}_{\nu}$, but rather $P^{-\mu}_{\nu}$. Thus the general solution 
can be written as 
\begin{equation}
R_l(t,t') = A_l(t') (1-t^2)^{-\frac{1}{4}} P_{\frac{1}{2}}^{l + \frac{1}{2}}(t) + B_l(t') (1-t^2)^{-\frac{1}{4}} P_{\frac{1}{2}}^{-l - \frac{1}{2}}(t) \, ,
\end{equation}
which is Eq.\eqref{dgf.Req3}.

%{From the general expression for Legendre functions of the second kind defined between $(-1,1)$ \cite{MO2:1966}  
%		
%\begin{equation}
%\frac{2}{\pi} \sin (\pi \mu) Q_{\nu}^{\mu}(x) = \cos (\pi \mu) P_{\nu}^{\mu}(x) - \frac{\Gamma(\nu + %\mu + 1)}{\Gamma(\nu - \mu + 1)} P_{\nu}^{- \mu}(x)\, ,
%\end{equation}
%
%we see that $Q^{\mu}_{\nu} =0$ when $\nu =\frac{1}{2}$ and $\mu = l + \frac{1}{2}$, with $l=1,2,\cdots$. In this case, $Q^{\mu}_{\nu}$ is not the other independent solution, but rather $P^{-\mu}_{\nu}$ (which is verified through the non-vanishing Wronskian $W(P^{\mu}_{\nu},P^{-\mu}_{\nu})$). 
%
%
%This leads to the general solution in this case being given by
%
%
%\begin{equation}
%R_l(t,t') = A_l(t') (1-t^2)^{-\frac{1}{4}} P_{\frac{1}{2}}^{l + \frac{1}{2}}(t) + B_l(t') %(1-t^2)^{-\frac{1}{4}} P_{\frac{1}{2}}^{-l - \frac{1}{2}}(t) \, ,
%\end{equation}}}
%

%%%%%%%%%%%%%%%%%%%%%%%%%%%%%%%%%%%%
\section{Limits of the de Sitter solutions as $y \to 1$} \label{app2}
%%%%%%%%%%%%%%%%%%%%%%%%%%%%%%%%%%%
Let us first note that the hypergeometric functions given in Eq.\eqref{dgf.Rsol} 
are of the form $\, _2 F_1 \left( a, a+1 ; 2 a + \frac{3}{2} ; y^2 \right)$\,, 
where $a = \frac{l}{2}$ and $a = \frac{ - l - 1}{2}$ correspond to the two 
hypergeometric functions contained in $g_l(y)$ and $f_l(y)$ respectively.  
There exists a known formula for evaluating the hypergeometric functions 
at the point $y^2 = 1$. This formula is given by (cf. Eq.~(15.4.20) of~\cite{NIST})
\begin{equation}
\, _2 F_1\left(a, b, c, 1 \right) = \frac{\Gamma\left(c\right)\Gamma\left(c - a - b\right)}
{\Gamma\left(c - a\right) \Gamma\left(c - b\right)} \qquad  
\Re\left(a + b - c\right) < 0\,; \;c \neq 0 , -1, -2, \dots \, 
\label{app2.lim}
\end{equation} 
This formula applies to the hypergeometric functions included in $f_l$ and $g_l$, 
but not to their derivatives. Let us consider the functions separately to find 
their derivatives at $y^2 = 1\,.$
%The $g_l$ hypergeometric functions and their derivatives can be determined through 
%the integral representation for the hypergeometric functions.
%
%%%%%%%%%%%%%%%%%%%%%%%%%%%%%%%%%%%%%%%%%%%%
\subsection{$f_l(y)$ solutions and hypergeometric functions}
%%%%%%%%%%%%%%%%%%%%%%%%%%%%%%%%%%%%%%%%%%
For the $f_l$ solutions, we need to find the expressions explicitly in order to 
determine the nature of the derivatives at the point $y = 1$. For the values of 
$l=0, 1, 2$ and $3$, the corresponding hypergeometric functions are, respectively,
\begin{align}
_2 F_1 \left(-\frac{1}{2}, \frac{1}{2}; \frac{1}{2}; y^2\right) \qquad\qquad & (l=0) \,; \notag\\
%\qquad  \qquad 
\, _2 F_1 \left(-1, 0; -\frac{1}{2}; y^2\right) \qquad\qquad &  (l=1) \,;\notag\\
_2 F_1 \left(- \frac{3}{2}, -\frac{1}{2}; - \frac{3}{2}; y^2\right) \qquad\qquad & (l=2) \,; \notag\\
\qquad  \qquad  \, _2 F_1 \left(-2, -1; -\frac{5}{2}; y^2\right) \qquad\qquad &  (l=3)\,.
\label{app2.4sols}
\end{align}
Here we see that $\, _2 F_1 \left(a, b; c; y^2\right)$ and $\, _2 F_1 \left(a-1, b-1; c-2; y^2\right)$  
represent two successive even (odd) solutions when $\left(a, b; c\right) = \, 
\left(\frac{-l-1}{2}, \frac{-l+1}{2}; - l + \frac{1}{2}\right)$\,.

Two hypergeometric functions which are contiguous are related to one another 
through certain differentiation formulas (cf. Eq.s~(15.5.4) and (15.5.9) of~\cite{NIST}). 
Let us look at the
ones relevant to the $f_l$ functions. These are
\begin{align}
%\, _2 F_1 \left(a+n, b+n; c+n; z \right) &= \frac{ (c)_n}{(a)_n (b)_n} \frac{d^n}{dz^n} \,_2 F_1 \left(a, b; c; z \right) \, ,\notag \\ 
%\, _2 F_1 \left(a, b; c+n; z \right) &=  \frac{ (c)_n}{(c - a)_n (c - b)_n} \left(1 - z \right)^{c + n - a - b} \frac{d^n}{dz^n} \left[ \left(1 - z \right)^{a + b - c}\,_2 F_1 \left(a, b; c; z \right) \right] \, . \label{app2.gsol}\\
%~&~ \notag \\
\, _2 F_1 \left(a-n, b-n; c-n; z \right) &= \frac{1}{(c-n)_n}(1-z)^{c+n-a-b} z^{1+n-c} 
\frac{d^n}{dz^n}\,\times \notag \\
&\qquad \qquad \times\, \left[ \left(1 - z \right)^{a + b - c} z^{c-1}\,_2 F_1 \left(a, b; c; z \right) \right] \, ,\notag\\
 \, _2 F_1 \left(a, b; c-n; z \right) &= \frac{1}{(c-n)_n} z^{1+n-c} \frac{d^n}{dz^n} 
 \left[ z^{c-1}\,_2 F_1 \left(a, b; c; z \right) \right] \, ,
\label{app2.fsol}
\end{align}
where $(k)_n = \frac{\Gamma \left(k + n\right)}{\Gamma \left( k \right)}$ is Pochhammer's symbol. 
Using the two relations in Eq.\eqref{app2.fsol}, we can write 
%
%\begin{align}
%\, _2 F_1 \left(a+n, \right. & \left. b+n; c+ 2 n; z \right) \notag\\
%& = \frac{ (c + n)_n (c)_n}{(c-a)_n (c-b)_n (a)_n (b)_n} (1-z)^{c - a -b} \, \frac{d^n}{dz^n} \left[ (1 - z)^{a+b+n-c}\frac{d^n}{dz^n} \,_2 F_1 \left(a, b; c; z \right) \right] \, .
%\label{app2.g1}
%\end{align}
%
%Similarly, using the equations in Eq.\eqref{app2.fsol}, we can construct the following equation
%
\begin{align}
\, _2 F_1 \left(a-n,  \right.  &\left. b-n; c- 2 n; z \right) \notag\\
& =  \frac{z^{1 + 2n - c}}{(c - n)_n (c - 2 n)_n} \frac{d^n}{dz^n} 
 \left[(1 - z)^{c+n-a-b} \right. \, \times \notag \\
 & \qquad \qquad \qquad\left. \times \, 
\frac{d^n}{dz^n}  \left[z ^{c-1} (1 - z)^{a+b-c} \, _2 F_1 \left(a, b; c; z \right) \right] \right]
\label{app2.f1g}
\end{align}

Using $\, _2 F_1 \left(a, b; c; z \right) = \, _2 F_1 \left(a, a+1; 2a + \frac{3}{2}; z \right)$ in Eq.\eqref{app2.f1g} provides the relevant recursive relation for the $f_l$ 
hypergeometric functions. To simplify the 
notation in what follows, let us define
%%, those for which $a = -\frac{l+1}{2}$ in 
%$\, _2 F_1 \left(a, a+1; 2a + \frac{3}{2}; z \right)$.
%
\begin{equation}
\, _2 F_1 \left(\frac{-l - 1}{2}, \frac{-l + 1}{2}; - l + \frac{1}{2}; z\right) = F_l(z) \, .
\label{app2.rec}
\end{equation}
With this definition, the solutions we seek are given by $f_l(y) = 
\frac{F_l(y^2)}{y^{l+1}}$. Combining Eq.\eqref{app2.rec} with 
Eq.\eqref{app2.f1g}, we can write
%
%
%\begin{align}
%%\, _2 F_1 & \left(a+n. a+n+1; 2 a + 2 n + \frac{3}{2}; z \right) \notag\\
%%& = \frac{ (2a + \frac{3}{2} + n)_n (2a + \frac{3}{2})_n}{(a + \frac{3}{2})_n (a + 1)_n (a + \frac{1}{2})_n (a)_n } (1-z)^{\frac{1}{2}} \frac{d^n}{dz^n} \left[ (1 - z)^{n - \frac{1}{2}}\frac{d^n}{dz^n} \,_2 F_1 \left(a, a+1; 2a + \frac{3}{2}; z \right) \right] \, ,\label{app2.g}\\
%\, _2 F_1 & \left(a-n,  a + 1- n; 2a- 2 n + \frac{3}{2}; z \right) \notag\\
% & =  \frac{z^{2n - 2a - \frac{1}{2}}}{(2a + \frac{3}{2} - n)_n (2a + \frac{3}{2} - 2 n)_n} \frac{d^n}{dz^n} \left[(1 - z)^{n+ \frac{1}{2}} \frac{d^n}{dz^n} \left[z ^{2a + \frac{1}{2}} (1 - z)^{-\frac{1}{2}} \, _2 F_1 \left(a, a+1; 2a + \frac{3}{2}; z \right) \right] \right] \, .
%\label{app2.f}
%\end{align} .
%
\begin{align}
F_{l + 2n}(z) = & \frac{z^{2n + \frac{1}{2} + l}}
{\left(l + \left(n -\frac{1}{2}\right)\right)_n \left(l + \left(2 n -\frac{1}{2}\right)\right)_n} \times \notag \\ & \qquad \quad \times
\frac{d^n}{dz^n} \left[(1 - z)^{n+ \frac{1}{2}} 
\frac{d^n}{dz^n} \left[z ^{- l - \frac{1}{2}} (1 - z)^{-\frac{1}{2}} F_l(z) \right] \right]\,.
\label{app2.f}
\end{align}
%
%Eq.\eqref{app2.f} reveals that every even and odd $l$ solution can be derived 
%from the lowest $l$ solution. More relevant in determining the general form of 
%the functions for even and odd $l$, is the fact that these functions can be determined 
%iteratively by setting $n=1$ and making a different choice of $l$ during each iteration. 
%We will now elaborate on how this approach can be used to find the general expressions.
%%  We will need this equation to find the exact expressions for \emph{any} $f_l(y)$ solution of interest, which will be elaborated on in the following subsection. An analagous relation to the one provided in Eq.\eqref{app2.f} can also be derived for the $g_l$ hypergeometric functions, this time relating $\, _2 F_1 \left(a+n,  a + 1+ n; 2a + \frac{3}{2} + 2n; z \right)$ and $\, _2 F_1 \left(a,  a + 1; 2a + \frac{3}{2}; z \right)$. We haven't described them here, since as we described earlier, it will be far simpler to use the integral representation for the hypergeometric functions to determine the limits of the $g_l$ hypergeometric functions and their derivatives at the point $y=1$. 
For the recursion relation, we need only consider Eq.\eqref{app2.f} with $n=1\,,$
\begin{equation}
F_{l + 2}(z) = \frac{z^{\frac{5}{2} + l}}{\left(l + \frac{1}{2}\right) 
	\left(l + \frac{3}{2}\right)} \frac{d}{dz} \left[(1 - z)^{\frac{3}{2}} 
\frac{d}{dz} \left[z ^{- l - \frac{1}{2}} (1 - z)^{-\frac{1}{2}} F_l(z) \right] \right] \, ,
\label{app2.f1}
\end{equation}
which upon evaluating the derivatives can be written as
\begin{align}
 F_{l + 2}(z) =& \left(1 - \frac{(2l+1)(2l+2) z}{3 + 8l + 4l^2} \right)F_l(z) \notag \\
 & \qquad\qquad - \frac{(8l + 4) z - (8l+2)z^2}{3 + 8l + 4l^2} F'_l(z) 
 - \frac{4(z^3 - z^2)}{3 + 8l + 4l^2}F''_l(z) \, ,
\label{app2.fodd}
\end{align}
where primes denote differentiation with respect to $z$. As we will see below, 
for even $l$ we can extract a factor of $\sqrt{1-z}$ to write the functions 
$F_l(z)$ in the form $F_l(z) = \sqrt{1 - z} D_l(z)$. Substituting this in  
Eq.\eqref{app2.f1}, we find a recursion relation for the functions $D_l(z)\,,$
\begin{align}
D_{l + 2}(z) =& \left[ \left(1 - \frac{2l (2l+1) z}{3 + 8l + 4l^2} \right)D_l(z) \right.
\notag \\
& \qquad\qquad \left.- \frac{(8l + 4) z - (8l-2)z^2}{3 + 8l + 4l^2} D'_l(z) 
- \frac{4(z^3 - z^2)}{3 + 8l + 4l^2}D''_l(z)\right]\,.
\label{app2.fev}
\end{align}

We will now need the lowest order solutions to proceed further. The lowest order 
solution for even $l$ is $F_0(z)\,,$ %%%$f_0(z)$,
 which corresponds to 
$ _2F_1 \left(-\frac{1}{2}, \frac{1}{2}; \frac{1}{2}; z \right)$\,,
as shown in Eq.\eqref{app2.4sols}. 
The hypergeometric function $\, _2 F_1 
\left(\frac{\alpha}{2}, -\frac{\alpha}{2}; \frac{1}{2}; z \right)$ has the 
following known representation
\begin{equation}
\, _2 F_1 \left(\frac{\alpha}{2}, -\frac{\alpha}{2}; \frac{1}{2}; z \right) 
= \cos \left( \alpha \sin^{-1} \left(\sqrt{z} \right) \right) \, ,
\end{equation}
and the case where $\alpha = -1$ is the one we require. 
The lowest order solution for odd $l$ is 
$f_1(z)$\,, which corresponds to $F_1(z) = \ _2 F_1 \left(-1, 0; 
-\frac{1}{2}; z \right)$. From the 
definition of the hypergeometric function $\ _2 F_1 \left(a, b; c; z \right)$ 
\begin{equation}
\ _2 F_1 \left(a, b; c; z \right) = \sum_{n=0}^{\infty} \frac{(a)_n (b)_n}{(c)_n} \frac{z^n}{n!} \, ,
\end{equation}
we know that $\ _2 F_1 \left(0, b; c; z \right) = \ _2 F_1 \left(a, 0; c; z \right) = \ _2 F_1 \left(a, b; c; 0 \right) = 1$. This tells us that the lowest order solutions are simply
\begin{align}
F_0(z) = \ _2 F_1 \left(-\frac{1}{2}, \frac{1}{2}; \frac{1}{2}; z \right) &= \sqrt{1-z} \, , \notag \\
F_1(z) = \ _2 F_1 \left(-1, 0; -\frac{1}{2}; z \right) &= 1\,.
\end{align}

We will use these to derive the expressions for the $f_l(y)$ functions 
given in Eq.\eqref{dgf.rep}. 
We begin with the even $l$ solutions. It can be seen that all even $l$ solutions 
are of the form $F_l(z) = \sqrt{1-z} D_l(z)$.  This follows directly from the fact that  
$F_0(z) =  \sqrt{1-z} D_0(z)$, where $D_0(z) = 1\,,$ and Eq.\eqref{app2.fev}. 
The operator in Eq.\eqref{app2.fev} takes a polynomial and produces another 
polynomial of one order higher. 
The only exception is $D_0(z)=1\,$ which, when inserted into 
Eq.\eqref{app2.fev}, produces $D_2(z) = 1\,.$ We can also calculate
 directly that
 $F_2(z) = \sqrt{1-z} = \sqrt{1-z}D_2(z)$. 
It follows from Eq.\eqref{app2.fev} that $D_{2k}(z)$ is a polynomial of order 
$(k-1)$ in $z$\,.

For $F_l(z)$ corresponding to odd $l$, we can use the recursion relation of Eq.\eqref{app2.fodd} directly.
%solutions follow in a similar fashion. 
For example, substitution of $F_1(z) = 1$ in Eq.\eqref{app2.fodd}
leads to $F_3(z) = 1 - \frac{4}{5} z$, substituting this result back in 
Eq.\eqref{app2.fodd} results in $F_5(z) = 1 - \frac{4}{3} z + 
\frac{8}{21} z^2$\,, etc. Thus $F_{2k+1}(z)$ is a polynomial of order $k$ in $z$\,.
%% We thus notice that odd $l$ solutions are given by $F_l(z)$, which 
%% is a linear polynomial in $z$ of order $\frac{l - 1}{2}$. 

We can now make a change of variable to $z=y^2$ in all the $F_l(z)$ solutions, 
and consider $f_l(y) = \frac{F_l(y^2)}{y^{l+1}}$ to find the 
$f_l(y)$ solutions shown in Eq.\eqref{dgf.rep}.

%%%%%%%%%%%%%%%%%%%%%%%%%%%%%%%%%%%%%%%%%%%%%%%%%%
\subsection{Limits of the $g_l(y)$ solutions}\label{appgl}
%%%%%%%%%%%%%%%%%%%%%%%%%%%%%%%%%%%%%%%%%%%%%%%%%%
The limit of the $g_l$ hypergeometric functions and their derivatives 
are most easily determined by making use of the following integral 
representation for hypergeometric functions (cf. Eq.~(15.6.1) of~\cite{NIST})
\begin{align}
\, _2 F_1\left(a, b, c, x \right) = \frac{\Gamma\left(c \right)}
{\Gamma\left(b \right) \Gamma\left(c - b \right)} &\int_{0}^{1} t^{b - 1} 
(1 - t)^{c - b - 1} (1 - x t)^{-a} dt \notag\\ & \qquad \qquad \left( \Re (c) > \Re (b) > 0 \right) \, ,
\label{app2.int}
\end{align}
As we have seen, the $f_l$ hypergeometric functions have $c \leq b$ for 
all choices of $l$, with the equality holding true for the $l=0$ case.
Hence, we could not use Eq.\eqref{app2.int} for those functions, and 
had to make use of the treatment described earlier. For the $g_l$ 
hypergeometric functions, $a= \frac{l}{2}$, which guarantees that 
$\, _2 F_1\left(a, a+1, 2a + \frac{3}{2}, y^2 \right)$ always has 
$c>b>0$. This also holds true for the derivative of this function 
on account of Eq.\eqref{dgf.hypr}.  We can thus consider Eq.\eqref{app2.int} 
in terms of the $g_l$ hypergeometric functions we are dealing with, 
in which case we have
\begin{equation}
\, _2 F_1\left(\frac{l}{2},\frac{l}{2}+1, l + \frac{3}{2}, y^2 \right) = \frac{\Gamma\left( l + \frac{3}{2} \right)}{\Gamma\left(\frac{l}{2} + 1\right) \Gamma\left(\frac{l+1}{2} \right)} \int_{0}^{1} \left(\frac{t}{1 - t y^2}\right)^{\frac{l}{2}} (1 - t)^{\frac{l - 1}{2}} dt  \, ,
\label{app2.int2}
\end{equation}
while the derivative of this function takes the form
\begin{equation}
\partial_y\left(\, _2 F_1\left(\frac{l}{2}, \frac{l}{2}+1, l + \frac{3}{2}, y^2 \right)\right) = \frac{\Gamma\left(l + \frac{3}{2} \right) l y }{\Gamma\left(\frac{l}{2} + 1\right) \Gamma\left( \frac{l + 1}{2} \right)} \int_{0}^{1} \left(\frac{t}{1 -t y^2}\right)^{\frac{l}{2}+1} (1 - t)^{\frac{l - 1}{2}} dt  \, ,
\label{app2.intd}
\end{equation}

The explicit representation of the $g_l$ hypergeometric functions and their derivatives,  in terms of elementary functions, can now be derived using these equations for specific choices of $l$. Since we are interested in the nature of the limit of these functions as $y \to 1$ for any choice of $l$, we can simply take this limit in the above expressions, and then evaluate the integrals. This amounts to the evaluation of standard integrals. We find that Eq.\eqref{app2.int} gives us the following finite result
\begin{equation}
\, _2 F_1\left(\frac{l}{2},\frac{l}{2}+1, l + \frac{3}{2}, 1\right) = \frac{ 2 \sqrt{\pi}}{(l+1)} \frac{\Gamma \left(l + \frac{3}{2} \right)}{\left(\Gamma \left(\frac{l + 1}{2} \right)\right)^2} \, ,
\end{equation} 
while Eq.\eqref{app2.intd} diverges for all choices of $l$. 

Similarly, the $y \to 0$  limits can also be determined very simply by using substituting $ y=0$ in  Eq.\eqref{app2.int2} and (\ref{app2.intd}. One can easily find that $\, _2 F_1\left(\frac{l}{2},\frac{l}{2}+1, l + \frac{3}{2}, 0\right) = 1$, by working out the integral. The derivative of this function at $y=0$ vanishes on account of an overall factor of $y$, and the fact that the integral is finite. 

%%%%%%%%%%%%%%%%%%%%%%%%%%%%%%%%%%%%%%%%%%%%%%%%% 

\end{document}